\documentclass{lhep}  

\usepackage{amsmath}
\usepackage{amssymb}
\usepackage{braket}
\usepackage{slashed}     

\journal{LHEP}
\vol{xx}
\jyear{2018}
\pages{xxx} 

\received{xx January 2018}
\published{xx March 2018}

\def\be{\begin{equation}}
\def\ee{\end{equation}}
\def\bea{\begin{eqnarray}}
\def\eea{\end{eqnarray}}

\def\La{\mathcal{L}}

\newcommand{\nn}{\nonumber}

\title{Neutrino oscillation caused by spacetime geometry}
\author{Indrajit Ghose, Riya Barick, Amitabha Lahiri}
\address{S.~N.~Bose National Centre for Basic Sciences, JD Block, Sector - III, Salt Lake, Kolkata - 700106, West Bengal, INDIA}

\begin{document}

\begin{abstract}
Effects of gravitational interaction are generally neglected in particle physics. A first order formulation of gravity is presented to include gravity in Quantum Mechanical Lagrangian of fermions. It is seen that fermions minimally coupled to gravity gives rise to a torsionless effective theory with a quartic interaction. After passing through a thermal background the most generic form of contortion contributes to neutrino effective mass. This effective mass can change the current oscillation parameters.
\end{abstract}

\maketitle

\section{Introduction}
\label{sec:intro}

In the Standard Model, neutrinos are the neutral partners of the charged leptons of each family. As neutrinos do not have a vertex in the unbroken $U(1)$ gauge group part they only interact by massive vector boson exchange. As a result it is very difficult to detect neutrinos even though the flux of solar neutrinos at the Earth's surface is on the order of $10^{11} cm^{-2} s^{-1}$~\cite{scientific_american}. After its detection in 1956~\cite{reins_cowan}, Ray Davis Jr. in 1968~\cite{Davis:1968cp,Cleveland:1998nv} tried detecting solar neutrinos following the radiochemical method proposed by Pontecorvo~\cite{Pontecorvo:1946mv}. This led to the solar neutrino anomaly~\cite{bahcall_davis}. It is now widely accepted that Pontecorvo's idea of flavor mixing is the explanation of discrepancy between observed and predicted solar neutrino flux.

The idea of neutrino mixing by Pontecorvo is quite simple -- the time evolution operator is not diagonal in the production-detection basis of neutrinos. In the sun the neutrinos are produced as flavor eigenstates and the detector also can only detect the flavor eigenstates. However, as the neutrino propagate through space a linear combination of the flavor eigenstates named as mass eigenstates evolve as stationary states. The relation between the two different set of basis vectors are given by a mixing matrix. They are related as~\cite{ParticleDataGroup:2020ssz}
\be
\ket{\nu_{\alpha}}=\sum_{i}U^{*}_{\alpha i}\ket{\nu_{i}}\,. \label{eq:mixing_matrix}
\ee
The $\ket{\nu_{i}}$'s give the mass basis and $\ket{\nu_{\alpha}}$'s are the flavor basis.

In the present article we will study the effect of gravity on neutrinos\footnote{Based on talk delivered by Indrajit Ghose at the International Conference on Neutrinos and Dark Matter (NuDM-2022), Sharm El-Sheikh, Egypt, 25-28 September, 2022.}. 
The article will be arranged as follows. In Sec.~\ref{sec:gravity} we discuss the dynamics of fermions in curved space-time, leading to an effective four-fermion interaction. Using this we find an additional contribution in the background averaged Lagrangian for neutrinos propagating through matter, in Sec.~\ref{sec:background_neutrino}. Finally we use this to calculate the refraction probability of neutrinos in flavor space in Sec.~\ref{sec:nu_osc}.

\section{Fermions under gravity}
\label{sec:gravity}

Effects of gravitational interaction is usually neglected in particle physics. However, in scenarios of astroparticle physics the effects may not be negligible. Let us consider how to write a diffeomorphism invariant actions for fermions  on curved spacetime. Usually Quantum Field Theory is defined on a flat Minkowskian manifold. To construct a theory invariant under general coordinate transformation thus requires defining the theory on the Minkowskian tangent manifold at each point and soldering that to the curved spacetime. In 3+1 dimensional spacetime, this can be done via tetrads or vierbeins, defined by
\be
g=e^{T}\eta e\,, \label{eq:tetrad_def}
\ee
or in index notation
\be
g_{\mu\nu}=e_{\mu}^{i}\eta_{ij}e^{j}_{\nu}\,, \label{eq:tetrad_def_index}
\ee 
where $g$ is the spacetime metric and $\eta$ is the Minkowskian metric. In order to show explicitly that the two kinds of indices live on two different spaces we denote the spacetime indices by Greek letters and tangent space indices by Latin letters. The inverse tetrad $e^{\mu}_{i}$, also called the cotetrad,  is defined by 
\be
e^{\mu}_{i} g_{\mu\nu}e^{\nu}_{j} = \eta_{ij}
\ee
The covariant derivative is taken to be tetrad-compatible (thus metric compatible), which results in the relation sometimes referred to as the tetrad postulate, 
\be
	e^\lambda_i \partial_\mu e^i_\nu + 	A_{\mu}{}^{i}{}_{j} e^j_\nu e^\lambda_i - \Gamma^\lambda{}_{\mu \nu} = 0\,.
\ee
$A_{\mu}{}^{i}{}_{j}$ are components of the spin connection and $\Gamma^\lambda{}_{\mu \nu}$ are the spacetime connection components.
{In general, the $\Gamma^\lambda{}_{\mu \nu}$ are not assumed to be symmetric in the lower indices -- the antisymmetric part corresponds to torsion.
} This enables us to write the Ricci sclar in terms of tetrads and spin connection, 
\be
R=e_{i}^{\mu}e_{j}^{\nu}(\partial_{[\mu}A_{\nu]}{}^{ij}+A_{[\mu|}{}^{i}{}_{k}A_{|\nu]}{}^{kj})\,. \label{eq:Ricci_scalar}
\ee
The spin connection and can be thought of as the gauge potential arising from the local Lorentz invariance of Quantum Field Theory in Curved Spacetime.

The action for free fermions on curved spacetime is
\be
S=\int d^{4}x |e|(\frac{1}{2\kappa}R + \sum_{f}\bar{f}(i\slashed{D}-m_f)f)\,. \label{eq:fermion_under_gravity} 
\ee
with the spinor covariant derivative being defined as~\cite{Kibble:1961ba,Sciama:1964wt,Hehl:1976kj,Hehl:1974cn,Hammond:2002rm,Hehl:2007bn,Gasperini:2013}
\be
D_{\mu}\psi=\partial_{\mu}\psi-\frac{i}{4}A_{\mu}{}^{ab}\sigma_{ab}\psi\,, \label{covder}
\ee
where $\sigma_{ab}=i/2[\gamma_{a},\gamma_{b}]$\,. Then the action in Eq.~\eqref{eq:fermion_under_gravity} is diffeomorphism invariant. This is the theory of fermions minimally coupled to gravity through the spin connection. 
We can vary the action with respect to the spin connection and get the equation of motion, which has the solution
\be
A_{\mu}{}^{ab}=\omega_{\mu}{}^{ab}+ \frac{\kappa}{8}\sum_{f}\bar{\psi}_f\{\gamma_{c},\sigma^{ab}\}e^{c}_{\mu}\psi_f\,.
\label{connection.1}
\ee 
{Here $\omega_{\mu}{}^{ab}$ corresponds to the usual symmetric Levi-Civita connection, while the second term is what is known as contortion $\Lambda_{\mu}{}^{ab}$\,, generated fully by the fermion fields. This corresponds to a torsion $T_{\mu\nu\rho} = \Lambda_{\mu ab}e^a_\nu e^b_\rho\,,$ which is completely antisymmetric in all of its indices. For the solution of Eq.~(\ref{connection.1}) we can write $T_{\mu\nu\rho} = \epsilon_{\mu\nu\rho\lambda}\sum_{f}\bar{\psi}_f \gamma_{a}\gamma^5 e^{a\lambda}\psi_f$\,.  Thus the fermions are coupled to the torsion as part of their coupling to the spin connection. However, the symmetries of the Lagrangian also allow a more general form of the contortion.} 

It was recently proposed that the most general form of the contortion is~\cite{Chakrabarty:2019cau}
\be
\label{chiral.torsion}
	\Lambda_{\mu}{}^{ab} = \frac{\kappa}{4}\epsilon^{abcd}e_{c\mu} \sum\limits_f \left(\lambda^f_{L}\bar{f}_{L}\gamma_d f_{L} + \lambda^f_{R}\bar{f}_{R}\gamma_d f_{R}\right)\,.
\ee
Here the $\lambda$'s are coupling parameters and the sum runs over all species of fermions. Plugging in this expression of the contortion and rescaling the $\lambda$'s to absorb a factor of $\kappa$\,,  we get the fermion part of the Lagrangian as
\begin{align}
\La&=\frac{1}{2\kappa}R+ \sum_{f}\bar{f}(i\slashed{\hat{D}} -m_f)f-\sum_{f}(\bar{f}\gamma^{\mu}(\lambda_{f}+\lambda'_{f}\gamma^{5})f)^2\,, \label{eq:neutrino_lagrangian}
\end{align}
in which $\hat{D}_{\mu}$ denotes the contortion-free spinor covariant derivative. The coupling constants $\lambda$ now have mass dimension $-1$. The four-fermion interaction appearing here causes a modification of neutrino oscillations.


The central question in neutrino oscillation concerns the probability that neutrinos produced in one flavor will be detected later in the same flavor. In flat space, the effective Lagrangian for Dirac neutrinos interacting via weak interactions can be written as~\cite{Wolfenstein:1977ue}
\begin{align}
\La_{\nu} =& \sum_{i}\bar{\nu}_{i}(i\slashed{\partial} - m_i)\nu_{i} \nn \\ 
&-\sum_{\alpha}\sum_{f}\frac{G_F}{\sqrt{2}}\bar{f}\gamma^{\mu}(g_1-g_2\gamma^{5})f\bar{\nu}_{\alpha}\gamma_{\mu}(1-\gamma^{5})\nu_{\alpha}\,.
\nn \\ 
&-\frac{G_F}{\sqrt{2}}\bar{e}\gamma^{\mu}(1-\gamma^{5})e\bar{\nu}_e\gamma_{\mu}(1-\gamma^{5})\nu_e\,. 
\label{eq:neutrino_wolfenstein}
\end{align}
Here $i$ denotes the mass basis of neutrinos, $\alpha$ is the flavor index, and $f$ denotes every other species of fermion. Also, according to standard model $g_1=T_{3}^{f}-2Q^{f}\sin^2\theta_W$ and $g_2=T_{3}^{f}$~\cite{Weinberg:1967tq}. $Q^{f}$ is the charge of fermion $f$\,, $T_{3}^{f}$ is the third component of isospin, $\theta_W$ is the weak mixing angle.
If we now include the four-fermion interaction arising from the neutrinos being on curved spacetime, we get
\begin{align}
\La_{\nu} =& \sum_{i}\bar{\nu}_{i}(i\slashed{\hat{D}} - m_i)\nu_{i} -\sum_{i}(\lambda_i\bar{\nu}_{i}\gamma^\mu\mathbb{L}\nu_i)\sum_f(\bar{f}\gamma_{\mu}(\lambda_{f}+\lambda'_{f}\gamma^{5})f) \nn \\
&-\sum_{\alpha}\sum_{f}\frac{G_F}{\sqrt{2}}\bar{f}\gamma^{\mu}(g_1-g_2\gamma^{5})f\bar{\nu}_{\alpha}\gamma_{\mu}(1-\gamma^{5})\nu_{\alpha}\,
\nn \\ 
&-\frac{G_F}{\sqrt{2}}\bar{e}\gamma^{\mu}(1-\gamma^{5})e\bar{\nu}_e\gamma_{\mu}(1-\gamma^{5})\nu_e \,.
\label{eq:neutrino_lagrangian_gravity}
\end{align}
For a normal matter distribution like earth and sun the curvature is not very large. Hence the Levi-Civita covariant derivative $\hat{D}_{\mu}$ can be approximated by ordinary partial derivatives. Then $e^{ip\cdot x}$ are again the free particle states and the standard techniques of Quantum Field Theory are applicable.  Here, we have suppressed the neutrino-neutrino self interaction as it will not be relevant in our calculations. Eq.~\eqref{eq:neutrino_lagrangian_gravity} gives the Lagrangian governing neutrinos, with $\hat{D}_\mu$ replaced by $\partial_\mu$ in regions of small curvature.

\section{Neutrinos through a medium}
\label{sec:background_neutrino}

Neutrinos propagating through a medium undergo numerous collisions with the thermalised background. $S$ matrix elements for neutrinos must be calculated using thermal field theory. There are different formulations of Finite Temperature Field Theory~\cite{Bellac:2011kqa,Niemi:1983nf,Kapusta:2006pm,Das:1997gg,Das:2000ft}, we will use the real time formalism developed by Schwinger as this is convenient for our calculations. In this formalism the fermion propagator will be changed~\cite{Pal:1989xs}\,,
\be
D(p)=\frac{i(\slashed{p}+m)}{p^2-m^2+i\epsilon}-2\pi n_{f}(p^{0})\delta(p^2-m^2)(\slashed{p}+m)\,.
\ee
Here $n_f$ is the Fermi-Dirac distribution for the fermion of mass $m$. Due to the presence of the momentum conserving delta function, the propagator is still a Green function of the Dirac equation.\footnote{The authors thank Prof.~P.~B.~Pal for making this clear.} In the following discussion we will assume that the temperature is not so high that the gauge boson propagators also receive thermal corrections, which is true in the case of atmospheric and solar neutrinos. Let us now calculate the background averaged Lagrangian after propagation through a neutral medium.

At first let us consider the charged current mediated processes. The corresponding Feynman diagram is {($l_\alpha$ is the charged lepton corresponding to $\nu_\alpha$)}\vskip-0.2cm
	\begin{figure}[h]
		\begin{center}
					\includegraphics{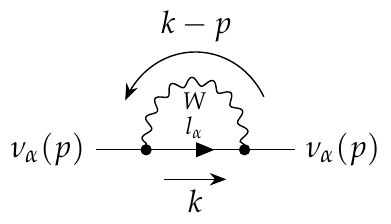}
		\end{center}
		\end{figure}
\vskip-0.7cm
This diagram will change the neutrino propagator as
\begin{align}
\frac{i}{\slashed{p}}(-i\Sigma)\frac{i}{\slashed{p}}\,.
\end{align}
A straightforward calculation yields
\begin{align}
-i\Sigma&=-\frac{iG_F}{\sqrt{2}}\int \frac{d^4k}{(2\pi)^4}\gamma^{\lambda}(1-\gamma^5)\,\times \nn \\
&[\frac{i(\slashed{k}+m)}{k^2-m^2+i\epsilon}-2\pi n_{f}(k^{0})\delta(k^2-m^2)(\slashed{k}+m)]\gamma_{\lambda}(1-\gamma^5) \nonumber
.\end{align}

The first term in the parenthesis yields the counter term for the mass renormalisation. This will be infinite and we are not interested in that. Let us consider that we are working with an already renormalised theory. Then the second term yields
\begin{align}
-i\Sigma=&-\frac{iG_F}{\sqrt{2}}\int \frac{d^4k}{(2\pi)^3 2\omega_k}\gamma^{\lambda}(1-\gamma^5)(\slashed{k}+m)\gamma_{\lambda}(1-\gamma^5) \nn \\
&\quad [\delta(k^0-\omega_k)+\delta(k^0+\omega_k)] n_{f}(k^0) \nn \\
&=-i\sqrt{2}G_Fn_e\gamma^0\mathbb{L}\,. \label{eq:cc_amplitude}
\end{align}
We have used the definition
\begin{equation}
n_e=2\int \frac{d^3k}{(2\pi)^3}n^e_f(\omega_k)\,,
\end{equation}
which is the total electron density. Then the dressed propagator is 
\begin{align}
&\frac{i}{\slashed{p}}+\frac{i}{\slashed{p}}(-i\Sigma)\frac{i}{\slashed{p}} \nn \\
=&\frac{i}{\slashed{p}-\Sigma}=\frac{i}{\gamma^{0}(p^{0}-\sqrt{2}G_Fn_e\mathbb{L})-\vec{\gamma}\cdot\vec{p}}\,.
\end{align}
Let us now consider the neutral current mediated process. The corresponding Feynman diagram is (where $f$ is any fermion present in the background)
	\begin{figure}[h]
		\begin{center}
		\includegraphics{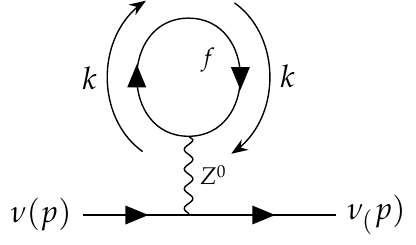}
	\end{center}
	\end{figure}
\vskip-0.7cm
A calculation similar to the previous one shows that due to the neutral current mediated process the propagator is modified as~\cite{Pal:1989xs}
\begin{align}
&\frac{i}{\slashed{p}}+\frac{i}{\slashed{p}}(-i\Sigma)\frac{i}{\slashed{p}} \nonumber \\
=&\frac{i}{\slashed{p}-\Sigma} \nonumber \\
=&\frac{i}{\gamma^0(p^0+\sqrt{2}G_F\frac{n_n}{2}\mathbb{L})-\vec{\gamma}\cdot \vec{p}}\,,
\end{align}
where $n_n$ is the total neutron density. Hence the total change in propagator pole in the electron neutrino is
\be
\frac{i}{\gamma^0(p^0-(\sqrt{2}G_Fn_e\mathbb{L}-\sqrt{2}G_F\frac{n_n}{2}\mathbb{L}))-\vec{\gamma}\cdot \vec{p}}\,.
\ee
This is the familiar result originally derived by Wolfenstein~\cite{Wolfenstein:1977ue}. 

We can similarly calculate the effective matter potential for the torsional four-fermion interaction. The corresponding Feynman diagram is 
	\begin{figure}[h]
		\begin{center}
					\includegraphics{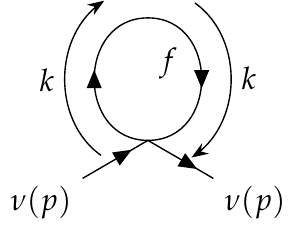}
		\end{center}
		\end{figure}

Following the same method as in the case of the neutral current mediated process, we can evaluate the shift in the pole of the propagator of type $i$ neutrino as
\begin{align}
\frac{i}{\gamma^{0}(p^0-\sum_f \lambda_fn_f\lambda_i\mathbb{L})-\vec{\gamma}\cdot\vec{p}}=\frac{i}{\gamma^{0}(p^0-\lambda_i\tilde{n}\mathbb{L})-\vec{\gamma}\cdot\vec{p}}~~,
\end{align}
where we have defined $\tilde{n}=\sum_f \lambda_fn_f\,,$ the sum running over background fermions, i.e., protons, neutrons, and electrons. 

These calculations enable us to write
\begin{align}
\La_{\nu}=&\sum_{i}\bar{\nu}_{i}(i\slashed{\partial}-m_i)\nu_{i}-\tilde{n}\sum_{i}\lambda_{i}\nu_{i}^{\dagger}\mathbb{L}\nu_{i}\nn \\
&
-\sqrt{2}G_Fn_e\nu_e^{\dagger}\mathbb{L}\nu_e+G_F \frac{n_n}{\sqrt2}\sum_\alpha\nu_{\alpha}^{\dagger}\mathbb{L}\nu_{\alpha} 
\,. \label{eq:neutrino_averaged_lagrangian_with_gravity}
\end{align}

\section{Neutrino oscillation}
\label{sec:nu_osc}

Experimentally it is observed that $\sin\theta_{13}$ is very small. As a result, a two flavor model is often a very good starting point. We can analyze the solar neutrino problem by considering oscillations between the electron- and muon-type neutrinos. The atmospheric neutrino can also be analyzed by considering muon neutrino to tau neutrino oscillations. Let us therefore briefly discuss the two flavor model of neutrino oscillations.

From Eq.~\eqref{eq:neutrino_averaged_lagrangian_with_gravity} we can write~{\cite{pal-mohapatra, kim}}
\begin{align}
	i\begin{pmatrix}{\dot\nu_e} \\ {\dot\nu_{\mu}}\end{pmatrix} =\left[E_0 +\frac{1}{4E}\begin{pmatrix}-\Delta m_s^2\cos 2\theta +D & \Delta m_s^2\sin 2\theta\\ \Delta m_s^2 \sin 2\theta & \Delta m_s^2 \cos2\theta - D\end{pmatrix}\right]
	\begin{pmatrix}{\nu_e} \\ {\nu_{\mu}}\end{pmatrix} ,
	\label{eq:TDSE_for_2nu_2}
\end{align}
where we have written $D = 2\sqrt{2}G_F n_e E\,,$ defined 
\begin{equation}
	E_0 = E+\frac{m_1^2+m_2^2}{4E}+\frac{\lambda_1+\lambda_2}{2}\tilde{n} -\frac{G_F}{\sqrt{2}}(n_n-n_e)\,,
\end{equation}
and also defined

\begin{equation}\label{ms-squared}
		\Delta m_s^2 =\Delta m^2+2 \tilde{n}E \Delta \lambda\,,
\end{equation}
{where} $\Delta m^2=m_2^2-m_1^2\,$ and $\Delta\lambda = \lambda_2 - \lambda_1\,.$
Let us write $\theta_M$ for the mixing angle in matter, modified by the torsional interaction,
\begin{equation}
	\tan 2\theta_M=\frac{\tan 2\theta}{1-\frac{D}{\Delta m_s^2 \cos 2\theta}}\,.
\end{equation}
Then we can diagonalize Eq.~(\ref{eq:TDSE_for_2nu_2}). 
The eigenvalues are $E_0 \mp\frac{\Delta m_M^2}{4E}\,,$ resulting in the survival probability
\begin{equation}
	P_{\nu_e \to \nu_e}=1-\sin ^2(2\theta_M)\sin^2\left(\frac{\Delta m_M^2}{4E}L\right)\, \label{eq:survival_probability_in_matter_for_2nu}
\end{equation}
and the conversion probability 
\begin{equation}
	P_{\nu_e \to \nu_{\mu}}=\sin ^2(2\theta_M)\sin^2\left(\frac{\Delta m_M^2}{4E}L\right)\,,
	 \label{eq:conversion_probability_in_matter_for_2nu}
\end{equation}
Where we have written 
\be
\Delta m_M^2=\sqrt{(\Delta m_s^2\cos 2\theta-D)^2+(\Delta m_s^2 \sin 2\theta)^2}\,.
\ee

{Eqs.~(\ref{eq:survival_probability_in_matter_for_2nu}) and (\ref{eq:conversion_probability_in_matter_for_2nu}) are the simplest examples of neutrino oscillation being affected by the four-fermion interactions induced by spacetime geometry. The $\nu_e\to \nu_\mu$ conversion probability becomes relevant for solar neutrinos, for which however a varying matter density should be considered. Probabilities of conversion and survival for other flavors of neutrino become relevant in other processes. Then the geometrical coupling constants can be found, along with the bare masses $m_i$ of the neutrinos, by fitting the respective formulae to the experimental data. We mention in passing that in principle there could be oscillation among neutrino flavors even when $m_i=0$ provided the coupling constants $\lambda_i$ are not all equal. }

\section{Conclusion and Remarks}
\label{sec:conclu}

We have used Einstein-Cartan theory to describe fermions under gravity. In this framework the torsion couples to the spin of fermions, the most general form of the  coupling being a combination of vector and axial currents. The vector part of the contortion contributes to the refractive index of the medium at the lowest order. One remarkable feature of this model is that now there is a part of the refractive index of the medium which survives even if neutrinos are massless or have quasi-degenerate spectra. This means massless or quasi-degenrate neutrinos have a non-vanishing conversion probability, which turns out to be independent of the energy of the incident neutrinos. Our results are expected to affect estimates of the oscillation parameters. 

{One might think that the coupling constants $\lambda\,,$ being of mass dimension $1/M$\,, should be suppressed by $M_{Pl}$\,, which is the natural mass scale of quantum gravity. But that would not be correct. In the case of torsion, the mass scale is not known.  Furthermore, the lack of a theory of quantum gravity also means that we do not know if torsion can be described by a renormalizable theory. Hence the dependence of the coupling parameters with 4-momentum is not known. Thus the couplings may not necessarily be suppressed by the Planck mass. In the current context, torsion is  generated by the fermions, leading to a four-fermion interaction, not determined by any other theory. Thus the couplings $\lambda$ can be fixed only by appealing to experimental data. 
}

This model can be easily extended to oscillations among three neutrino flavors. We can include the effects of torsion into the Hamiltonian for that as well and find the conversion probability in presence of torsion. Details will be presented elsewhere.


\end{document}